# Scaling the topological transport based on an effective Weyl model

Shen Zhang[1,2], Jinying Yang[1], Meng Lyu[1], Junyan Liu[1], Binbin Wang[1], Hongxiang Wei[1], Claudia Felser[3], Wenqing Zhang[4], Enke Liu[1,*], Baogen Shen[1,2]

[1]Beijing National Laboratory for Condensed Matter Physics, Institute of Physics, Chinese Academy of Sciences, Beijing 100190, China
[2]Ningbo Institute of Materials Technology & Engineering, Chinese Academy of Sciences, Ningbo, Zhejiang 315201, China
[3]Max Planck Institute for Chemical Physics of Solids, 01187 Dresden, Germany
[4] Department of Physics, Southern University of Science and Technology, Shenzhen 518055, China

*Email: ekliu@iphy.ac.cn

Magnetic topological semimetals are increasingly fueling interests in exotic electronic-thermal physics including thermoelectrics and spintronics. To control the transports of topological carriers in such materials becomes a central issue. However, the topological bands in real materials are normally intricate, leaving obstacles to understand the transports in a physically clear way. Parallel to the renowned effective two-band model in magnetic field scale for semiconductors, here, an effective Weyl-band model in temperature scale was developed with pure Weyl state and a few meaningful parameters for topological semimetals. Based on the model, a universal scaling was established and subsequently verified by reported experimental transports. The essential sign regularity of anomalous Hall and Nernst transports was revealed with connection to chiralities of Weyl nodes and carrier types. Upon a double-Weyl model, a concept of Berry-curvature ferrimagnetic structure, as an analogy to the real-space magnetic structure, was further proposed and well described the emerging sign reversal of Nernst thermoelectric transports in temperature scale. Our study offers a convenient tool for scaling the Weyl-fermion-related transport physics, and promotes the modulations and applications of magnetic topological materials in future topological quantum devices.

## I. INTRODUCTION

Magnetic topological semimetals are a class of materials that have attracted increasing attention in condensed matter physics in recent years [1-13]. In this class of materials, many novel physical behaviors have emerged in various fields[14-22]. Meanwhile, the magnetic topological materials also show potential applications in next-generation thermoelectric materials, spintronics, and quantum computing, thanks to the Berry curvature of topological bands[2,3,23-25]. To control the transports of topological carriers thus becomes a key issue to facilitate the practical applications of topological materials. However, the topological bands in most topological materials are relatively complex, and the electronic transport is frequently contributed by both dominative topological states and topologically trivial bands[14,26-29]. This case usually results in a disagreement between theoretical calculations and experimental analysis, which hinders the efficient applications of topological materials. The similar case has appeared in the fundamental research and semiconductor industry. A two-band model[30-33], which has been developed and widely used in semiconductors and metals, effectively reduces the band parameters that control the electronic transports (Fig. 1(a,b)). With the aid of two-band model, people can easily obtain a concise and clear picture for hole or electron carriers in semiconductors or spintronics, without probing the subtle band structures but just performing the convenient measurements on transverse normal Hall and longitudinal magnetoresistance as functions of magnetic fields. The concentration and mobility of carriers can be thus obtained by transport measurements. In current case of topological semimetal materials, to establish an effective topological band model, namely, a Weyl-band model, is also expected to simplify the electronic structures to pure Weyl bands and to bring intuitive physical picture as well as convenient property analysis in temperature scale (Fig. 1(c,d)). This will promote the practical applications of emergent magnetic topological materials in future topological magneto-electronics.

On the other hand, rich progresses have been made on the important topological transports including anomalous Hall effect (AHE) and anomalous Nernst effect (ANE) after the realization of magnetic topological semimetals[1-3,6,34-40]. The AHE and ANE induced by strong Berry curvature in magnetic topological materials show excellent performance over the traditional magnetic materials. Currently, however, many problems such as controlling the sign reversal and temperature dependence of these transports are unresolved, especially for their future applications in advanced thermoelectrics and spintronics. Recently, based on the doping effect in our previous reports on magnetic Weyl semimetal $Co_3Sn_2S_2$ [2,4,35], an experimental design of high-efficiency ANE thermoelectric thermopile device was realized by superimposing anomalous Nernst signals with different signs[41]. This study exhibits the great necessity and possibility of controlling the ANE signs for topological thermoelectrics by tuning the Fermi level of the materials. Thus, the regularity of signs and their underlying mechanism necessitate an in-depth study, which will facilitate the applications of topological transports in magnetic topological materials and their functional devices.

In this study, we propose an effective Weyl model that can describe the signs and temperature dependence of the AHE and ANE from the view of intrinsic mechanism. The reported data of magnetic topological materials also conform to the theoretical formula of this model. Then, we use the effective Weyl model to reveal the substratal sign regularity of AHE and ANE, and propose a picture of the Berry-curvature magnetic structure under multiple pairs of Weyl nodes to explain the emerging important phenomenon of sign reversal in ANE in magnetic Weyl semimetals $CoS_2$, $Co_3Sn_2S_2$[34], and $Fe_{3-\delta}GeTe_2$[42]. In this way, we can systematically categorize the properties of magnetic topological materials using the effective Weyl model, which is beneficial for the understanding and application of AHE and ANE effects in magnetic topological materials.

## II. METHODS

### A. Effective Weyl model and temperature scaling

$CoS_2$ single crystals were synthesized via the flux method in an $Al_2O_3$ crucible sealed within a quartz tube. The tube was heated to 800°C over 12 hours, maintained for 24 hours for sufficient reaction between sulfur and cobalt, and then slowly cooled to 600°C over 7 days for crystal growth. The crucible was then transferred from the muffle furnace at 600 °C to a centrifuge for flux removal.

### B. Electrical and thermal transport measurements

The measurements on were performed on the Physical Property Measurement System (PPMS) using a commercial thermal transport measuring rod and nanovoltmeters. The measured samples were cut out of a cuboid. During transport measurements, the magnetic field was perpendicular to the sample plane and the heat current was along long side.

### C. Thermal transport analysis

We measured the longitudinal and transverse voltages, the longitudinal and transverse temperature differences, and the heating power of the heater on the sample. After including the sample size, we obtained the longitudinal electric field intensity $E_x$, the transverse electric field intensity $E_y$, the longitudinal temperature gradient $\nabla_x T$, the transverse temperature gradient $\nabla_y T$, and the heat flux density $q_x$. Then, neglecting the secondary terms in the thermal transport equations, we obtained the Seebeck coefficient $S_{xx} = E_x / \nabla_x T$, Nernst coefficient $S_{yx} = E_y / \nabla_x T$, longitudinal thermoelectric conductivity $\alpha_{xx} = \sigma_{xx} S_{xx}$, transverse thermoelectric conductivity $\alpha_{yx} = \sigma_{yx} S_{xx} + \sigma_{xx} S_{yx}$, longitudinal thermal conductivity $\kappa_{xx} = -q_x / \nabla_x T$, and transverse thermal conductivity $\kappa_{yx} = -\kappa_{xx} \nabla_y T / \nabla_x T$.

## III. RESULTS AND DISCUSSION

### A. Effective Weyl model and temperature scaling

For the effective Weyl model, the simplest Weyl node energy band is adopted (Fig. 1(d)) to establish the single-Weyl model. Assuming a pair of Weyl nodes locate on the $z$-axis at coordinates $(0,0, \pm c)$, the Hamiltonian at $(0,0, -c)$ is $H = \lambda \boldsymbol{\sigma} \cdot \boldsymbol{k'} + \Delta$ ($\lambda > 0$), where $\sigma$ is the Pauli matrix and $\boldsymbol{k'} = (k_x, k_y, k_z + c)$. The corresponding energy dispersions of the upper and lower bands are $\varepsilon_\pm = \Delta \pm \lambda k'$, and the $z$-components of the Berry curvature from the upper and lower bands are $\Omega_\pm^z = \mp \frac{k'_z}{2k'^3}$. From the energy dispersion relation, one can infer that $\lambda$ in the Hamiltonian represents the slope of the linear bands, while $\Delta$ represents the relative deviation of the Weyl nodes from the Fermi level. Analyzing the $z$-component of the Berry curvature, one can see that the Berry curvature of the lower bands between this pair of Weyl nodes points in the positive $z$-direction, while the Berry curvature of the upper band points in the negative $z$-direction. Here, one only needs to consider the Berry curvature between $k_z = c$ and $k_z = -c$. Why not consider the Berry curvature outside $k_z = \pm c$? According to previous studies [43,44], under the influence of Landau levels, the band energy between a pair of Weyl nodes is lower than the outside band energy, and electrons tend to occupy on these bands rather than on bands outside the Weyl nodes.

According to the effective single-Weyl model, we can propose a quantitative description of the temperature dependence of the anomalous Hall conductivity $\sigma_{yx}^A$ and anomalous transverse thermoelectric conductivity $\alpha_{yx}^A$ based on the intrinsic picture for magnetic topological systems. The commonly used formula for intrinsic calculations [24,45] are

$$\sigma_{yx}^A = \frac{e^2}{\hbar}\int \frac{d\mathbf{k}}{(2\pi)^3} \Omega_z f, \qquad (1)$$

$$\alpha_{yx}^A = -\frac{ek_B}{\hbar}\int \frac{d\mathbf{k}}{(2\pi)^3} \Omega_z g. \qquad (2)$$

Here, $f = \frac{1}{1+e^{\varepsilon/k_B T}}$ represents the Fermi-Dirac distribution function and $g = \frac{\varepsilon f}{k_B T} + \ln(1 + e^{-\varepsilon/k_B T})$ represents another electron energy distribution function. The two equations satisfy the Mott relation [45,46] $\alpha_{yx}^A = -\frac{\pi^2}{3}\frac{k_B^2 T}{e}\frac{\partial \sigma_{yx}^A}{\partial \varepsilon_F}$ at low temperatures ($\varepsilon_F \gg k_B T$). In metals and semi-metals, the Fermi energy $\varepsilon_F > 1$ eV, while below 400 K the energy corresponds to the temperature $k_B T < 0.04$ eV. Thus, the Mott relation works under measurement conditions of 400 K and below. Based on the Mott relation, we can rewrite Eq. (2) as

$$\alpha_{yx}^A = \frac{\pi^2}{3}\frac{ek_B^2 T}{\hbar}\int \frac{d\mathbf{k}}{(2\pi)^3}\Omega_z \frac{\partial f}{\partial \varepsilon}. \qquad (3)$$

This equation is more advantageous for integration.

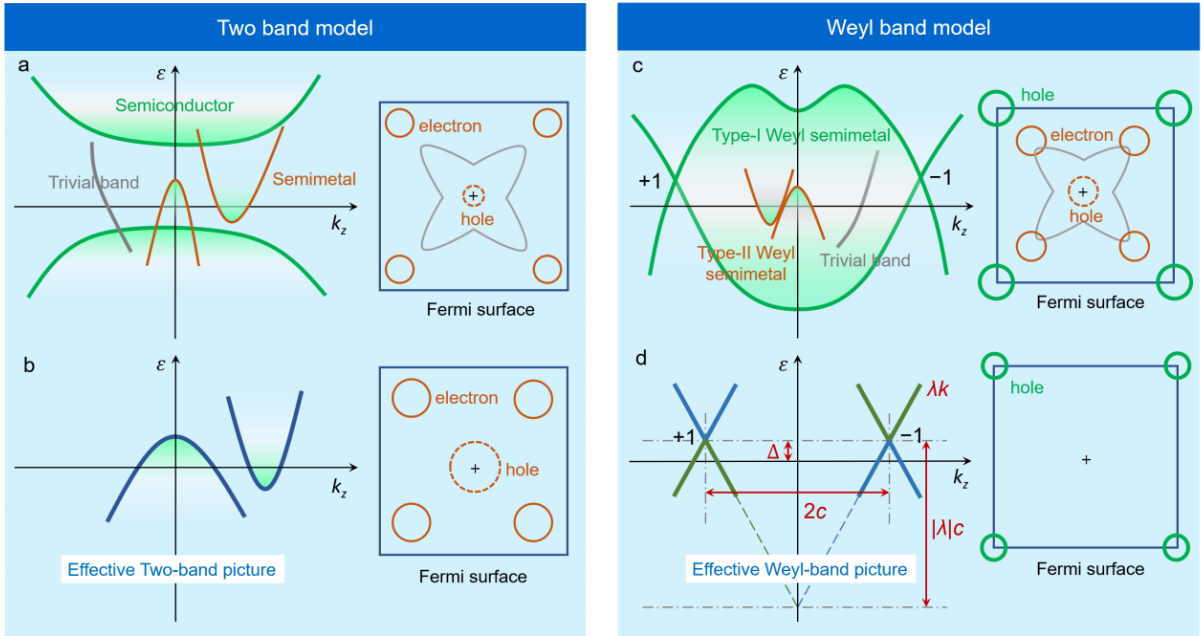

FIG. 1. Effective two-band model and effective Weyl-band model. (a) Multiple bands crossing the Fermi level contributing to transport in semimetals or dirty semiconductors. (b) Effective two-band model simplifying the bands to one or two parabolic bands. The influence of other bands is included in the carrier concentration and mobility of electrons and holes of effective two bands probed by resistivity measurements. (c) Band structures in topological materials are also not clean, with multiple types and multiple pairs of Weyl nodes as well as topologically trivial bands. Some bands with small energy gaps also contribute to the Berry curvature and play a role similar to Weyl nodes. (d) Effective Weyl band simplifying the bands in magnetic topological materials as one pair of Weyl nodes, considering only the linear dispersion region of these Weyl node bands. The parameters were reduced to describe the topological transports (such as AHE and ANE).

Then, we substitute the Berry curvature formula for the effective Weyl model into Eqs.(1)(3), and incorporate the influence of the magnetization [47] to obtain $\sigma_{yx}^A$ and $\alpha_{yx}^A$ in temperature scale. The unified temperature law for experimental behavior of collected magnetic topological materials can be well described according to our model (see the complete model in section 2 in Supplementary Material). For convenience, we further make an approximation of $\Delta \ll \lambda c$, that is, the distance from the Weyl node to the Fermi level is much smaller than the half of Weyl bandwidth, then the simplified formula of $\sigma_{yx}^A$ and $\alpha_{yx}^A$ can be respectively described by

$$\left|\sigma_{yx}^A\right| = \frac{M}{M_0} \frac{e^2}{8\pi^2 \hbar \lambda}(T_2 - T_1 - 2T\ln(1+e^{-T_1/T})), \tag{4}$$

$$\left|\alpha_{yx}^A\right| = \frac{M}{M_0} \frac{ek_B}{24\hbar\lambda} T\tanh(\frac{T_1}{2T}). \tag{5}$$

In these two equations, $T_1 = |\Delta|/k_B$ is related to the energy deviation of the Weyl nodes, and $T_2 = \lambda c/k_B$ is related to the bandwidth of the Weyl nodes. These two equations indicate that the temperature dependences of $\sigma_{yx}^A$ and $\alpha_{yx}^A$ depend on the magnetization and the thermal broadening of the electronic distribution function.

As shown in Fig. 2, the normalized master curve of our Weyl model can largely scale the experimentally reported temperature behavior of AHE and ANE of magnetic topological materials. The deviation of some data is due to simplification caused by approximation ($Fe_3Al$, $Mn_3Ge$). The deviation in simplified model indicates that the Weyl nodes to the Fermi level is comparable to the half of Weyl bandwidth ($\Delta \approx \lambda c$) in both systems; but the physics in both systems can be well captured by our complete model. The fitting based on the complete model can be found in the section 2 in Supplementary Material. Because all data of the anomalous Hall and anomalous Nernst effects are fitted by Eqs.(4)(5) separately, the parameters are not exactly consistent with each other. However, most of the fitting parameters are close to each other, which to large extent reflects the effectiveness of the model (see Tables S1 and S2).

By fitting the parameters, we can obtain the effective Weyl model formula corresponding to each material, and three parameters in the effective Weyl model are closely related to the magnitude of the anomalous Hall and anomalous Nernst effects. The farther the effective Weyl node is from the Fermi level, that is, the larger $\Delta$ is, the inflection point of the temperature-dependent $\sigma_{yx}^A$ and $\alpha_{yx}^A$ curves will appear at a higher temperature. This is because the farther the Weyl node is from the Fermi level, the stronger thermal perturbations are needed to affect its properties. The smaller the slope of the effective Weyl node, that is, the smaller $\lambda$ is, the larger $\sigma_{yx}^A$ and $\alpha_{yx}^A$ may be achieved, because the smaller the band slope, the more charge carriers may be involved in transport. The wider the bandwidth of the effective Weyl node, that is, the larger $\lambda c$ is, the larger $\sigma_{yx}^A$ may be achieved, because $\sigma_{yx}^A$ depends on the integral of the Berry curvature of the bands below the Fermi energy, and the wider the band width, the larger the range of integration energy.

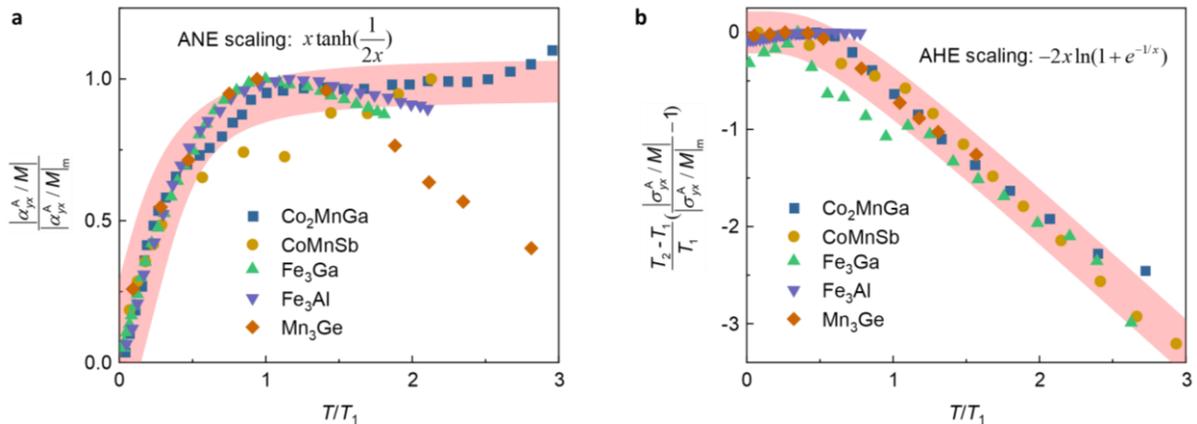

FIG. 2. Universal scaling of normalized temperature dependence of anomalous transverse transports for magnetic topological materials. Normalized temperature dependence of anomalous Nernst (a) and anomalous Hall (b) effects in magnetic topological materials. $T_1$ and $T_2$ are two parameters related to the effective Weyl nodes, and $x=T/T_1$.

## B. Sign regularity of AHE and ANE

The signs of AHE and ANE are extremely important for the applications of magnetic topological materials such as Nernst thermoelectrics. In different materials, there exist positive and negative values of anomalous Hall conductivity $\sigma_{yx}^A$ and anomalous transverse thermoelectric conductivity $\alpha_{yx}^A$, and some materials have also shown a change in sign of the ANE. We will use the effective Weyl model to explain the regularity of signs in different materials, and understand the behavior of sign reversal in the ANE.

The sign reversal behavior of ANE has been found in several magnetic topological systems (Fig. 3). We measured the single crystals of magnetic Weyl $CoS_2$ for its thermoelectric transport properties, including thermal conductivity, Seebeck coefficient, Nernst coefficient, and thermal Hall effect (see the section 1 in Supplementary Material). In the anomalous Nernst effect, we observed the phenomenon of sign reversal with temperature, with a maximal anomalous Nernst coefficient of 1.25 µV K$^{-1}$ and a maximal anomalous transverse thermoelectric conductivity of 7 A m$^{-1}$ K$^{-1}$. Similar sign reversals in ANE have also been observed in magnetic Weyl $Co_3Sn_2S_2$ [34] and magnetic topological $Fe_{3-\delta}GeTe_2$ [42]. In $Co_3Sn_2S_2$, the phenomenon of sign reversal was speculated to be related to the spin fluctuation and phonon, which is attributed to extrinsic scattering in the transport process. In $Fe_{3-\delta}GeTe_2$, a bulge in $\alpha_{yx}^A$ is observed near 20 K, which leads to a change in sign of $S_{yx}^A$ below 20 K. And the sign reversal is believed to originate from the contributions of the Fermi level and Berry curvature, which is attributed to the intrinsic band structure. In this work, the effective Weyl model can provide an intrinsic explanation for the sign reversal in these three materials.

To understand the sign regularity of intrinsic anomalous transverse transports, we first analogize it to the behavior of normal transverse transports. In normal Hall and normal Nernst behaviors, electrons moving in the longitudinal direction experience transverse deflection under the action of a magnetic field in the $z$ direction, and the sign of transverse current depends on the type of charge carrier. In the normal Hall effect, when the charge carrier is electron, the Hall conductivity $\sigma_{yx} > 0$, whereas when the charge carrier is hole, the Hall conductivity $\sigma_{yx} < 0$. In the normal Nernst effect, the transverse thermoelectric conductivity $\alpha_{yx} < 0$, regardless of whether the charge carrier is electron or hole.

The anomalous transverse transports are different from the normal transports. According to the intrinsic picture, the anomalous Hall conductivity $\sigma_{yx}^A$ and anomalous transverse thermoelectric conductivity $\alpha_{yx}^A$ in magnetic materials are determined by the Berry curvature in the $k$ space, which is also referred to as a pseudo-magnetic field [48]. Thus, anomalous transverse transport occurs under the action of the pseudo-magnetic field in the $z$ direction. The $z$-directional Berry curvature can either be aligned with or opposite to the external magnetic field, which can also affect the signs of $\sigma_{yx}^A$ and $\alpha_{yx}^A$. In a word, the sign of anomalous transverse transport is influenced by both the type of charge carrier and the direction of the Berry curvature.

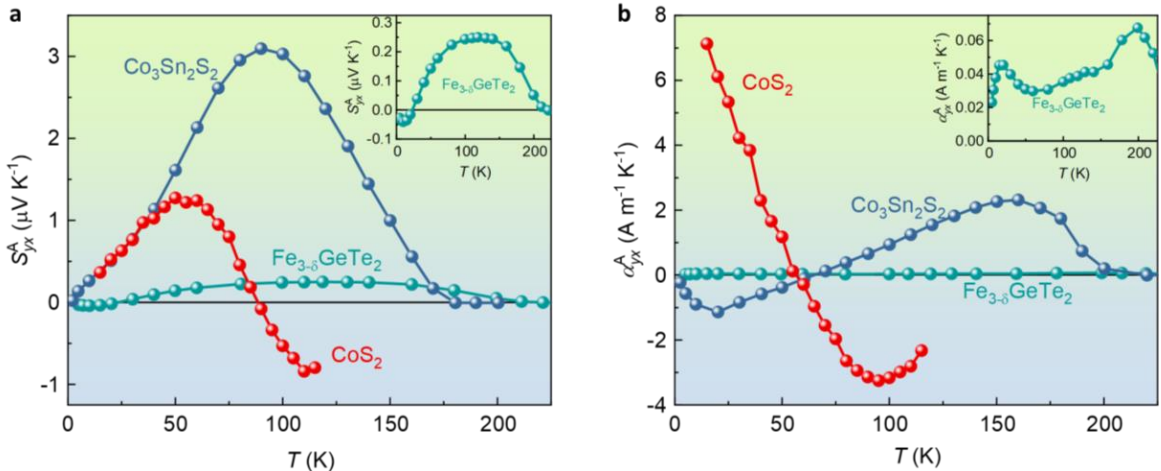

Fig. 3. Sign-change phenomenon of ANE in $CoS_2$, $Co_3Sn_2S_2$, and $Fe_{3-\delta}GeTe_2$. The temperature-dependent anomalous Nernst effect shows a sign-change phenomenon in $CoS_2$, $Co_3Sn_2S_2$, and $Fe_{3-\delta}GeTe_2$. In $CoS_2$ and $Co_3Sn_2S_2$, the phenomenon of a change in sign of $\alpha_{yx}^A$ with temperature has been observed. In $Fe_{3-\delta}GeTe_2$, a bulge in $\alpha_{yx}^A$ is observed near 20 K, which leads to a change in sign of $S_{yx}^A$ below 20 K.

The next question is how to determine the direction of the Berry curvature. Strong Berry curvature is often produced by topological states, such as Weyl nodes. The Weyl nodes are degenerate points of two bands, which can be referred to as the upper and lower bands. According to the definition of the Berry curvature [48], the magnitude of the Berry curvature of the upper and lower bands at the same $k$ point is equal, but their directions are opposite. Thus, does the Berry curvature exerting on charge carriers belong to the upper band or the lower band? Since charge carriers in transport mainly come from the vicinity of the Fermi surface, the Berry curvature felt by charge carriers is naturally the Berry curvature near the Fermi surface. When the Weyl nodes are above the Fermi level, the bands near the Fermi surface are the lower bands, and the Berry curvature felt by the charge carriers is from the lower bands. Similarly, when the Weyl nodes are below the Fermi level, the bands near the Fermi surface are the upper bands, and the Berry curvature felt by the charge carriers is from the upper bands. A subsequent question is, how do we know the directions of the Berry curvature for both upper bands and lower bands? According to the first-principles calculations [20], we can obtain the chirality of the Weyl nodes. A positive chirality corresponds to a diverging Berry curvature, and a negative chirality a converging Berry curvature. The direction of the Berry curvature revealed by the chirality refers to the direction of the Berry curvature of the lower bands at the Weyl nodes, while the direction of the Berry curvature from the upper bands is exactly opposite to the direction revealed by the chirality.

Then, we can analyze the sign of anomalous transverse transport induced by Weyl nodes in different cases (Fig. 4). In the effective Weyl model, the chirality of a pair of Weyl nodes can be represented as +1 and −1 respectively. In one chirality situation, the Weyl node with chirality +1 is located on the negative $z$-axis, while the Weyl node with chirality −1 on the positive $z$-axis (Fig. 4(a)), consistent with the situation in Fig. 1(d). In this chirality situation, when $\Delta > 0$, the Weyl nodes are above the Fermi level, and the carriers are holes. Holes feel the Berry curvature of the lower bands, which means that $\Omega_z > 0$. Treating the Berry curvature as a pseudo-magnetic field in $k$ space, according to the left-hand rule, we can obtain the sign of intrinsic anomalous transverse transport (Fig. 4(b)): $\sigma^A_{yx} < 0$, $\alpha^A_{yx} < 0$. When $\Delta < 0$, the Weyl nodes are below the Fermi level, and the carriers are electrons. Electrons feel the Berry curvature of the upper bands, which means that $\Omega_z < 0$. In this case, the sign of intrinsic anomalous transverse transport is as follows (Fig. 4(c)): $\sigma^A_{yx} < 0$, $\alpha^A_{yx} > 0$.

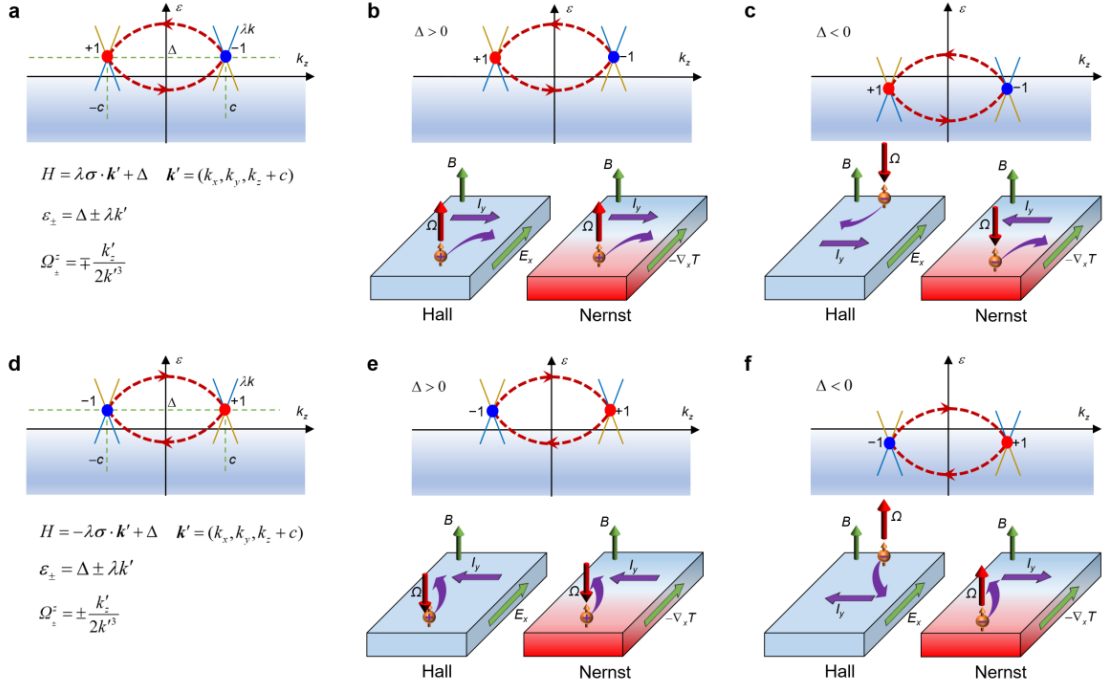

FIG. 4. Configurations of Weyl parameters for anomalous Hall and anomalous Nernst effects based on Weyl picture. Two cases of chirality of a pair of Weyl nodes: (a) positive chirality Weyl node along the negative $z$-axis and negative chirality Weyl node along the positive $z$-axis and (d) negative chirality Weyl node along the negative $z$-axis and positive chirality Weyl node along the positive $z$-axis. In both cases, the Weyl nodes can be above the Fermi level (b, e), or below the Fermi level (c, f). This results in four sign combinations for AHE and ANE (b, c, e, f). The sign of the AHE is only influenced by the chirality, while the sign of ANE is influenced by both the chirality and the energy position.

If we change the chirality of the Weyl nodes, let the Hamiltonian at $(0,0,-c)$ be $H = -\lambda\sigma \cdot \mathbf{k'} + \Delta$ (Fig. 4(d)), the energy dispersion of the upper and lower bands remains unchanged, and the $z$-component of the Berry curvature becomes $\Omega_\pm^z = \pm\frac{k'_z}{2k'^3}$. Now, the Berry curvature of the lower band points in the negative $z$-direction, while the Berry curvature of the upper band points in the positive $z$-direction. When $\Delta > 0$, we obtain $\sigma_{yx}^A > 0$, $\alpha_{yx}^A > 0$ (Fig. 4(e)). When $\Delta < 0$, then $\sigma_{yx}^A > 0$, $\alpha_{yx}^A < 0$ (Fig. 4(f)).

As a result, the sign of the anomalous Hall conductivity $\sigma_{yx}^A$ is determined solely by the chirality of the Weyl nodes, while the sign of the anomalous transverse thermoelectric conductivity $\alpha_{yx}^A$ is dominated by both the chirality of the Weyl nodes and the energy position of the Weyl nodes from the Fermi surface (corresponding to types of carriers). When modulating the Fermi level, causing the Weyl nodes to pass through the Fermi surface, the sign of $\sigma_{yx}^A$ remains unchanged, while the sign of $\alpha_{yx}^A$ reverses. This conclusion was partly confirmed by a previous study in which the role of energy position of Weyl has been addressed[49]. Here we provide a classical and intuitive figure and help to analyze the contribution of Berry curvature between multiple pairs of Weyl nodes.

### C. Double-Weyl model and Berry curvature structures

The sign reversal phenomenon was observed in the anomalous transverse thermoelectric conductivity $\alpha_{yx}^A$ with temperature in CoS$_2$ and Co$_3$Sn$_2$S$_2$ [34]. The experimental data cannot be fitted by Eq. (5), because this equation cannot deal with the sign reversal. To explain the experimental phenomenon, we further consider the contributions of multiple pairs of Weyl nodes to the anomalous transverse transport after examining the electronic structures. This is similar to the two-band model in semiconductors and metals[31,32], where the electrons and holes are taken into consideration to understand the non-linear field dependence of Hall resistivity and parabolic magnetoresistance. The contributions of multiple pairs of Weyl nodes to $\sigma_{yx}^A$ and $\alpha_{yx}^A$ can be considered independently of each other. The total $\sigma_{yx}^A$ and $\alpha_{yx}^A$ equal the sum of the contributions from each band, described by the equations

$$\sigma_{yx}^A = \frac{M}{M_0}\frac{e^2}{8\pi^2\hbar}\sum_{i=1}^{n}\frac{1}{\lambda_i}(T_{2i} - T_{1i} - 2T\ln(1+e^{-T_{1i}/T})), \tag{6}$$

$$\alpha_{yx}^A = -\frac{M}{M_0}\frac{ek_B T}{24\hbar}\sum_{i=1}^{n}\frac{1}{\lambda_i}\tanh(\frac{T_{1i}}{2T}), \tag{7}$$

where n represents the number of pairs of Weyl nodes, and $\lambda$ can be positive or negative to represent the chirality. To reduce the number of parameters used in fitting the experimental data, we only set n = 2, representing the contributions of multiple Weyl nodes through two pairs.

According to the density functional theoretical calculations of CoS$_2$, a total of 8 pairs of Weyl nodes was obtained, all above the Fermi level, but 2 pairs of Weyl nodes have opposite chiralities compared to the other 6 pairs [20] (Fig. 5(a)). Thus, we can classify these Weyl nodes into two groups, where the Berry curvature from Weyl nodes within the same group has the same direction, similar to a ferromagnetic alignment of magnetic moments, while the Berry curvature between the two groups has opposite directions, similar to an antiferromagnetic alignment. The effective double-Weyl picture can be shown in Fig. 5(b). Based on this model, we can separate the temperature-dependent $\sigma_{yx}^A$ and $\alpha_{yx}^A$ of CoS$_2$ into two independent contributions of two pairs of Weyl nodes with opposite polarized directions of Berry curvature (Fig. 5(c-f)). We also fit our theoretical equations with the experimental data of Co$_3$Sn$_2$S$_2$ and Fe$_{3-\delta}$GeTe$_2$ [34,42] and provide an effective explanation for the sign reversal in the anomalous transverse thermoelectric conductivity $\alpha_{yx}^A$ (see the section 3 in Supplementary Material).

From this view, the Berry curvature of multiple pairs of Weyl nodes can be analogized to the magnetic structures in real space of magnetic materials. Based on the direction and magnitude of each Berry curvature, the Berry-curvature magnetic structure can be likened to ferromagnetic, antiferromagnetic, ferrimagnetic-polarization structures, even more complex non-collinear magnetic structures. In this sense, the Berry curvature in single-Weyl model (Figs. 4) corresponds to the simple ferromagnetic structure. As we know, different magnetic structures lead to totally different thermo-magnetization curves of magnetic materials. Similarly, different Berry curvature structures would also lead to rich temperature dependences of anomalous transverse transports. This Berry curvature structure is applicable in more topological materials with symmetry breaking, even including the non-magnetic and non-centrosymmetric systems, which would promote further phenomena and effects in topological physics.

### IV. Conclusions

In summary, we proposed an effective Weyl model that lays the foundation for transport physics of topological semimetals, which provides general understanding of the temperature laws and their signs of anomalous Hall and Nernst effects. To explain the emerging sign reversal of anomalous Nernst effect, we propose that the Berry curvature of multiple pairs of Weyl nodes can be analogized to a real-space magnetic structure and provide corresponding theoretical formulas using the double-Weyl model. In real topological materials, there always are more than one pair of Weyl nodes. At the same time, Weyl nodes also exhibit various forms, such as type-I Weyl nodes, type-II Weyl nodes, nodal lines, and nodal planes, etc. In more general cases, the topological states and topologically trivial bands usually coexist in a system. Treating these complex topological states as well as the topologically trivial states in an effective Weyl band model could reduce the theoretical parameters and help to establish an intuitive understanding of measured properties of topological materials. On the one hand, the effective Weyl model can be used to analyse theoretical results and predict physical properties more clearly. On the other hand, the effective Weyl model is also expected to play a role in physical measurements other than electrical and thermal transports of topological spintronics and topological thermoelectrics.

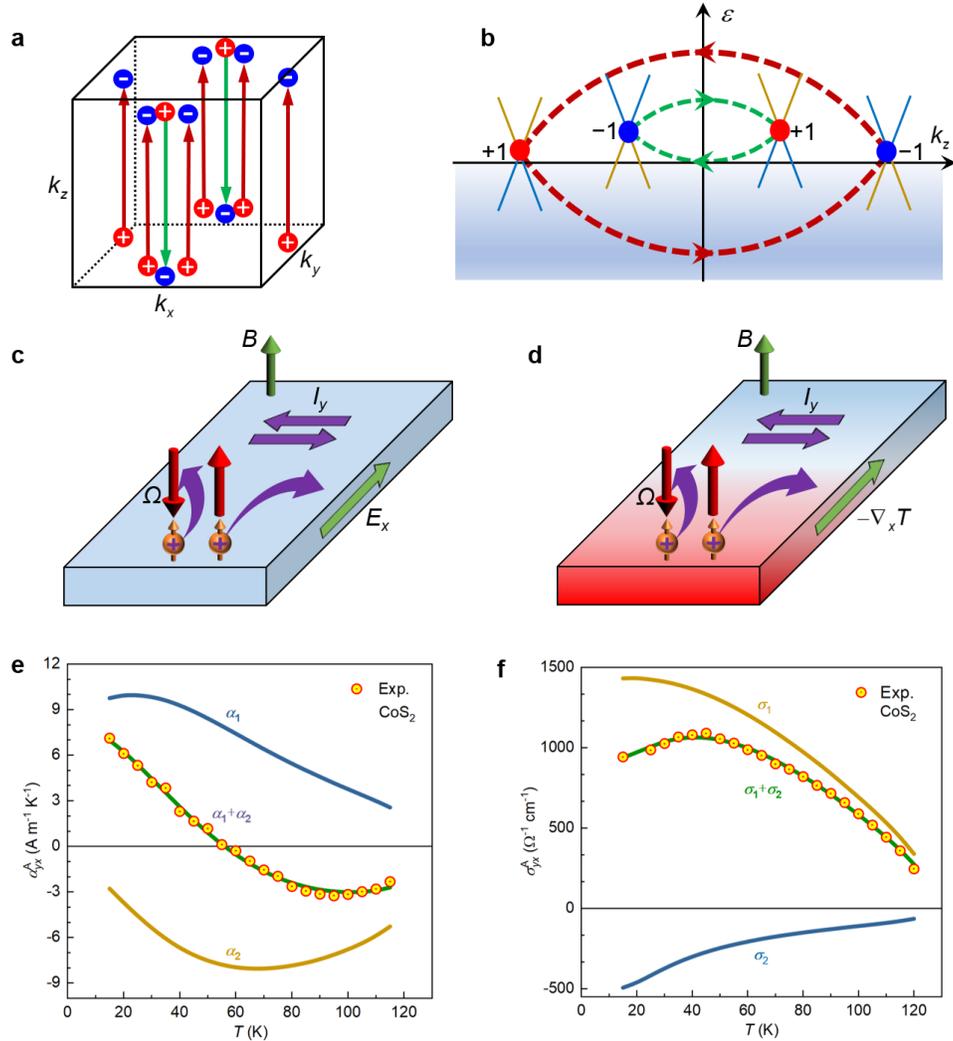

FIG. 5. Double-Weyl model proposed for the sign reversal of ANE in magnetic Weyl systems. (a) Positions of 8 pairs of Weyl nodes in the Brillouin zone of $CoS_2$ and the Berry curvature directions determined by chirality. (b) All Weyl nodes in $CoS_2$ are above the Fermi surface. Based on the Berry curvature direction, these 8 pairs of Weyl nodes can be classified into 2 groups, each represented by one pair of Weyl nodes. (c, d) Signs of the anomalous Hall and anomalous Nernst effects in $CoS_2$ under the double-Weyl model. (e, f) Anomalous Hall and anomalous Nernst curves are the sum of the contributions from two pairs of Weyl nodes in the double-Weyl model.

**Acknowledgments**

This work was supported by the National Key R&D Program of China (2019YFA0704900, 2022YFA1403800), the National Natural Science Foundation of China (Nos. 52088101, 12174426), the Strategic Priority Research Program (B) of the Chinese Academy of Sciences (CAS) (No. XDB33000000), the Synergetic Extreme Condition User Facility (SECUF), the CAS Project for Young Scientists in Basic Research YSBR-057, and the Scientific Instrument Developing Project of CAS (No. ZDKYYQ20210003).